\documentclass[4pt,twocolumn]{article}
\usepackage[sort &compress]{natbib}
\usepackage{graphicx}
\usepackage{color}
\usepackage[hidelinks]{hyperref}
\usepackage{skull}    
\usepackage{listings} 

\newcommand{\Webcolour}{\color{blue}}
\newcommand{\Link}[2]{\href{#1}{\Webcolour \bf #2}}
\usepackage[switch,columnwise]{lineno}

\newcommand{\target}{XZ~And}

\newcommand{\PR}[1]{{\color{magenta} {\ttfamily #1}}}

\newcommand{\RModel}[1]{{{model}$_{#1}$}}

\newcommand{\PM}{\!\!\pm\!\!}
\newcommand{\ON}{\!\!=\!\!}
\newcommand{\REJ}{$Q_F\!\!<\!\!10^{-16}$}
\newcommand{\ResultOne}{R1}
\newcommand{\ResultTwo}{R2}
\newcommand{\ResultThree}{R3}
\newcommand{\ResultFour}{R4}
\newcommand{\ResultFive}{R5}
\newcommand{\ResultSix}{R6}
\newcommand{\ResultSeven}{R7}
\newcommand{\ResultEight}{R8}

\newcommand{\threetest}{Test$^{3+3+3+3}$}
\newcommand{\fourtest}{Test$^{4+4+4}$}
\newcommand{\Dd}{\mathrm{[d]}}
\newcommand{\AD}{{^{*}}}
\newcommand{\HD}{{^{\dagger}}}
\newcommand{\WD}{{{^{\P}}^{9}}}
\newcommand{\RD}{{^{\bullet}}}
\newcommand{\JDdata}{JD-data}
\newcommand{\HJDdata}{HJD-data}
\newcommand{\Ddata}{D-data}
\newcommand{\Ysignal}{$1^{\mathrm{y}}$-signal}
\newcommand{\Ywindow}{$1^{\mathrm{y}}$-window}
\newcommand{\mdash}{~-~}
\newcommand{\pasp}{PASP}
\newcommand{\aap}{A\&A}
\newcommand{\aaps}{A\&AS}
\newcommand{\aj}{AJ}
\newcommand{\apj}{ApJ}

\newcommand{\apss}{Ap\&SS}
\newcommand{\mnras}{MNRAS}
\newcommand{\na}{NewA}

\title{About ten stars orbit eclipsing binary XZ Andromedae}
\author{{\bf Lauri Jetsu} \\
{\it Department of Physics, P.O. Box 64, FI-00014, 
University of Helsinki, Finland}; \\
email: lauri.jetsu@helsinki.fi}
\begin{document}
\twocolumn[ 
\begin{@twocolumnfalse}
\maketitle
\begin{abstract}
  A third body in an eclipsing binary system causes
  regular periodic changes in the 
  observed (O) minus the computed (C)
  eclipse epochs.
  Fourth bodies have rarely been
  detected from the O-C data.
  We apply the new Discrete Chi-square Method (DCM)
  to the O-C data of the eclipsing binary XZ Andromedae.
  These data contain the periodic signatures
  of at least ten wide orbit stars (WOSs).
  Their orbital periods are between 1.6 and 91.7 years.
  Since no changes have been observed in the
  eclipses of XZ And
  during the past 127 years,
  the orbits of all
  these WOSs are most probably co-planar.
  We give detailed instructions of how
  the professional and the amateur astronomers can easily
  repeat all stages of our DCM analysis with an ordinary PC,
  as well as apply this method to the O-C data of other
  eclipsing binaries.
  \\ ~ \\
  {{\bf Key words:} methods: data analysis - methods: numerical -
  methods: statistical - binaries: eclipsing \\ ~ \\}  
\end{abstract}
\end{@twocolumnfalse}
] 
  
\section{Introduction}

Naked eye observations
of Algol's eclipses
have been recorded into
the Ancient Egyptian 
Calendar of Lucky and Unlucky days
\citep{Por08,Jet13,Jet15,Por18}.
Today, 
eclipsing binary (EB) observations have
become routine for the professional
and the amateur astronomers.
On November 29th 1890,
Dr. Raymond S. Dugan
recorded the first primary eclipse
epoch of \target.
The last epoch
in our \target ~eclipse data
is from December 24th, 2017.
The primary (A4~IV-V, $3.2 m_{\odot}$, $2.4 R_{\odot}$)
and
the secondary (G~IV, $1.3 m_{\odot}$, $2.6 R_{\odot}$)
of this binary orbit each other
during $P_{\mathrm{orb}}=1.357$ days
\citep{Dem95}.

Periodic long-term changes are sometimes
observed O-C data of EBs.
A third or a fourth body can cause such
periodicity \citep[e.g.][]{Jet20B},
but there are also other alternatives,
like a magnetic activity cycle \citep[e.g.][]{App92}
or 
an apsidal motion \citep[e.g.][]{Bor05}.
Such long-term changes have also been observed
in \target ~\citep[][]{Dem95,Man16,Cha19}.
When \citet{Haj19} studied 
O-C data of 80~000 EBs,
they detected only 
four EB candidates that may have a fourth body.
Our preliminary analysis of \target ~with
the new Discrete Chi-Square Method (DCM)
already confirmed the presence of a third
and a fourth body
\citep[][]{Jet20}.
Here, we show that DCM can detect the periodic
O-C signals 
of at least ten bodies in this system.
Our appendix gives detailed instructions
for repeating every stage
of our DCM analysis.\footnote{
  DCM program code and all other necessary
  files are freely available
  in the 
  \Link{https://zenodo.org/record/3871549}{Zenodo
  database: doi {\color{red}10.5281/zenodo.3871549}}
All files, variables
  and other code
related items are 
printed in \PR{magenta} colour. 
}

\begin{figure*}
\begin{center}
  \resizebox{14.0cm}{!}
  {\includegraphics{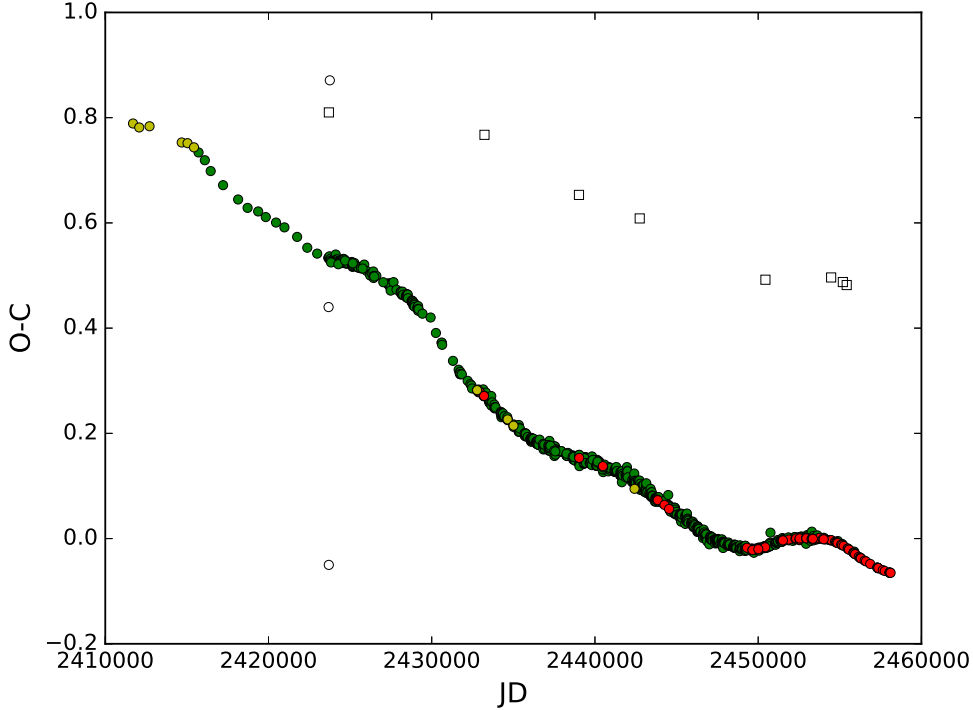}}
\end{center}
\caption{Original O-C data.
Analysed data 
(``C'' and ``E'' = red circles,
 ``V'' and ``F'' =  green circles,
``P'' and ``?'' = yellow circles)
and  rejected data
(secondary minima = transparent squares,
outliers = transparent circles).}
\label{FigOC}
\end{figure*}

\section{Data}

In September 2019,
we retrieved the observed (O) minus the computed (C)
primary eclipse epochs of \target  ~from the
\Link{https://www.bav-astro.eu/index.php/veroeffentlichungen/lichtenknecker-database/lkdb-b-r}
{Lichten\-knec\-ker-Database of the BAV}.
These data had been computed from the
ephemeris 
\begin{eqnarray}
\mathrm{HJD~}2452500.5129 + 1.35730911{\mathrm{E.}}
\label{EqEphe}
\end{eqnarray}
All original O-C data 
are shown in Fig. \ref{FigOC}.
We reject eight secondary minima
and
three primary minimum outliers
(Table \ref{TableRejected}).
The $n=1094$ 
Heliocentric Julian Day $t_i=t_{\odot,i}$ data
\begin{eqnarray}
y_i=y_{\mathrm{HJD}}(t_{\odot,i})={\mathrm{O-C}}
\label{EqOCHJD}
\end{eqnarray}
are our first analysed sample
(Table \ref{TableHJD}: hereafter \HJDdata).

The yearly distribution of data
is given in
Table \ref{TableMonths}.
Most of these observations
have been made between August and January.
The regular lack of observations
between February and July
is repeated in the data
for 127 years.
This ``\Ywindow''
can mislead DCM period analysis.
In our second analysed sample,
we create random time points
\begin{equation}
t_{i}^{\star}=t_{\odot,i}+ \delta t_i
\label{EqTstar}
\end{equation}
where $t_{\odot,i}$ time points are from Table \ref{TableHJD}.
The random shifts
$\delta t_i$ are uniformly distributed between
$- 365.^{\mathrm{d}}25/2$
and
$+ 365.^{\mathrm{d}}25/2$.
These new $t_i^{\star}$
time points are rearranged  into increasing
order.
No changes are made to $y_i$ and $\sigma_i$ values
of Table \ref{TableHJD}.
We analyse one such arbitrary sample of
$t_i^{\star}$, $y_i$ and $\sigma_i$
data (Table \ref{TableD}: hereafter
\Ddata).
We will show that the $\delta t_i$ random shifts
do not mislead the detection of longer
periods, but they do help us
in the identification of possibly
spurious periods
(Table \ref{Tabled13}).
For example, 
the largest possible
random shift of $\pm 183^{\mathrm{d}}$
is only about $\pm 6\%$
for $P_{\mathrm{min}}=3000^{\mathrm{d}}$,
which is our lower limit for
the tested longer periods of \target.

In our third sample,
we transform Table \ref{TableHJD} data to 
\begin{eqnarray}
  y_{\mathrm{JD}}(t_{\oplus,i})
  =y_{\mathrm{HJD}}(t_{\odot,i})-\delta t_i,
\label{EqOCJD}
\end{eqnarray}
where $\delta t_i = t_{\odot,i} - t_{\oplus,i}$,
and
$t_{\odot}$
and
$t_{\oplus,i}$
are the epochs when
the same eclipse of
\target ~is
observed in the Sun and on the Earth
(Table \ref{TableJD}: hereafter \JDdata).
No changes are made to $t_i$ and $\sigma_i$ values
of Table \ref{TableHJD}.
We use
\target ~coordinates 
$\alpha = 01^{\mathrm{h}}
~53^{\mathrm{m}} ~48.^{\mathrm{s}}76$
and $\delta = +41^{\mathrm{o}} ~51' ~24.97"$
in the transformation of Eq. \ref{EqOCJD}.
This transformation superimposes the
artificial ``\Ysignal'',
the Earth's motion around the Sun,
into these data.
We use this \Ysignal ~for checking
the reliability of our period analysis.
For \target, this effect is about $\pm 7.4$ minutes
$= \pm 0.0051$ days.

Since the errors $\sigma$ of the data
are unknown,
we use the following arbitrary relative weights
for different observations
\begin{itemize}
\item[] $w_1=3$ for ``C'' $(n=55)$ and ``E''$(n=29)$
\item[] $w_2=2$ for ``V'' $(n=939)$ and ``F'' $(n=61)$
\item[] $w_3=1$ for ``P'' $(n=7)$ and ``?'' $(n=3)$,
\end{itemize}
where the BAV observation systems in
German language are
``C = CCD'',
``E = Fotometer'',
``V = Visuel'',
``F = Fotoserie'',
``P = Platten''
and
``? = Unbekant''.
First, we fix the errors for the most
accurate ``C'' and ``E'' observations 
to $\sigma_1=0.^{\rm d}001$.
Then, the $w = \sigma^{-2}$ relation
gives the errors
$\sigma_2=\sqrt{2\sigma_1^2/3}=0.^{\rm d}00122$ 
for the ``V'' and ``F'' observations,
as well as
$\sigma_3=\sqrt{2\sigma_1^2/3}=0.^{\rm d}00173$
for the ``P'' and ``?'' observations.
The numerical values of these errors do
not influence our results, because we
use the same weight for every observation.
These errors are used only to give the correct
DCM code analysis 
format for our three files of data.


\section{Method}
\label{SectMethod}

The data are
$y_i=y(t_i) \pm \sigma_i$,
where $t_i$ are the observing
times and $\sigma_i$ are the errors
$(i=1,2, ..., n)$. 
The time span is $\Delta T=t_n-t_1$.
We apply DCM to these data.
This method can detect
many signals superimposed
on arbitrary trends.
Detailed instructions for
using the DCM python code
were already given in 
\citet[][Appendix]{Jet20}.
Here, we also provide all necessary
information for reproducing
our DCM analysis
of \target ~data.

DCM model is a sum of a periodic function $h(t)$
and an aperiodic function $p(t)$
\begin{eqnarray}
  g(t) = g(t,K_1,K_2,K_3)
   =  h(t) + p(t),
\label{Eqmodel}
\end{eqnarray}
where
\begin{eqnarray}
h(t)   & = & h(t,K_1,K_2) = \sum_{i=1}^{K_1} h_i(t) \label{Eqharmonicone} \\
h_i(t) & = & \sum_{j=1}^{K_2} 
B_{i,j} \cos{(2 \pi j f_i t)} + C_{i,j} \sin{(2 \pi j f_i t)}
\label{Eqharmonictwo} \\
p(t)   & = & p(t,K_3) = \sum_{k=0}^{K_3} p_k(t) \\
p_k(t) & =  & M_k \left[
                    {{2t} \over {\Delta T}}
                    \right]^k.
\label{Eqpolynomial}
\end{eqnarray}
The periodic $h(t)$ function is a sum of 
$K_1$ harmonic $f_i$ frequency signals $h_i(t)$.
The order of these signals is $K_2$. 
The trend $p(t)$ is an aperiodic $K_3$ 
order polynomial. The $g(t)$ model has
\begin{eqnarray}
p= K_1 \times (2K_2+1) + K_3+1
\end{eqnarray}
free parameters.
We use the abbreviation ``\RModel{K_1,K_2,K_3}''
for a model having orders
$K_1$, $K_2$ and $K_3$.
DCM determines the
$h_i(t)$ signal
parameters
\begin{itemize}

\item[] $P_i = 1/f_i = $ Period
\item[] $A_i = $ Peak to peak amplitude
\item[] $t_{\mathrm{i,min,1}} = $ Deeper primary minimum epoch 
\item[] $t_{\mathrm{i,min,2}} = $ Secondary minimum epoch (if present)
\item[] $t_{\mathrm{i,max,1}} = $ Higher primary maximum epoch
\item[] $t_{\mathrm{i,max,2}} = $ Secondary maximum epoch (if present),
\end{itemize}
and the $M_k$ parameters 
of the $p(t)$ trend.

We compute the DCM test statistic $z$ from the
sum of squared residuals $R$, because 
the errors for the data are unknown
\citep[][Eqs. 9 and 11]{Jet20}.
The $F=F_R$ test statistic gives
the Fisher-test critical levels $Q_F$
\citep[][Eqs. 13]{Jet20}.
When we compare a simple and a complex model,
the latter is a better model for the data if
\begin{eqnarray}
  Q_F < \gamma_F=0.001,
\label{EqFisher}
\end{eqnarray}
where 
$\gamma_F=0.001$ is the pre-assigned significance level
\citep[][Eq. 14]{Jet20}.

Our notations for the two signatures of unstable models
are

\begin{itemize}

  \item[] $\AD$ = Dispersing amplitudes
  \item[] $\HD$ = Intersecting frequencies
  
\end{itemize}
\noindent
The former can occur
without the latter,
but not vice versa \citep[][Sect. 4.3.]{Jet20}.
The periods that are clearly too large are
also denoted with $\HD$.
We use the notation

\begin{itemize}

  \item[] $\RD$ = Failed model

  \end{itemize}

  \noindent
  when $\AD$ ~and $\HD$ ~both occur,
  or the model must be rejected with
  the criterion of Eq. \ref{EqFisher}.
  
\section{Search for long periods}
\label{SectLong}

We first search for longer periods between
$P_{\mathrm{min}}=3000^{\mathrm{d}}$
and
$P_{\mathrm{max}}=100~000^{\mathrm{d}}$,
because  the \Ywindow ~and \Ysignal ~can mislead
the detection of periods below $P_{\mathrm{min}}$.
Later, we will also
search for periods
shorter than $P_{\mathrm{min}}$
(Sect. \ref{SectShort}).
Note that $P_{\mathrm{max}}> \Delta T=46~111^{\mathrm{d}}$,
because we will show that DCM can detect
periods longer than the time span of data.
We search for periodicity in
three different samples,
where the misleading
\Ywindow ~or \Ysignal ~is
present or absent.
These are

\begin{itemize}
\item[] \HJDdata: \Ywindow ~= Yes, \Ysignal ~= No
\item[] \Ddata: \Ywindow ~= No, \Ysignal ~= No
\item[]\JDdata: \Ywindow ~= Yes, \Ysignal ~= Yes
\end{itemize}

\begin{figure*}
\begin{center}
  \resizebox{12.5cm}{!}
  {\includegraphics{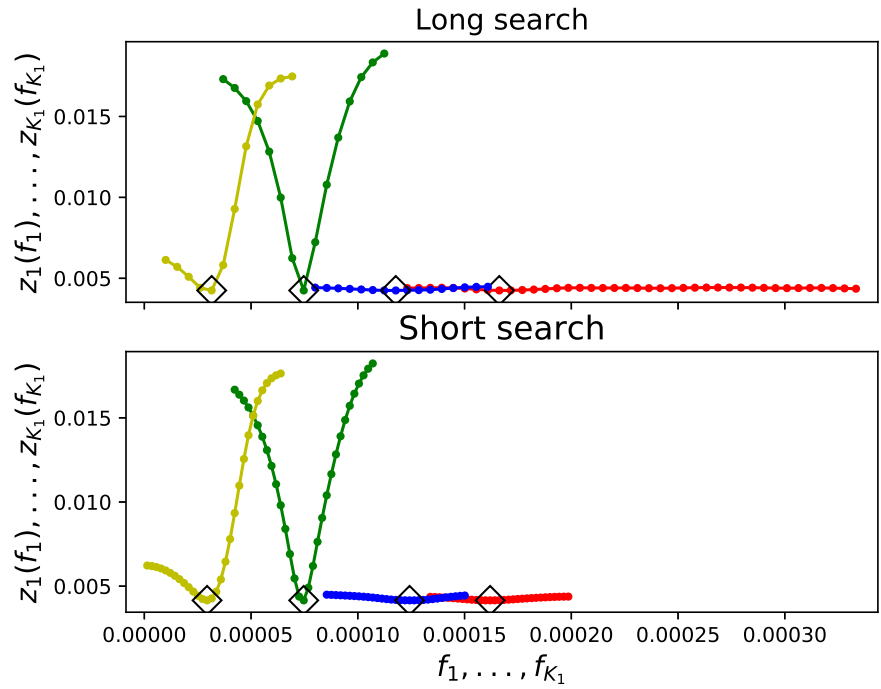}}
\end{center}
\caption{Periodograms for four signal \RModel{4,1,2} of
  Table \ref{TableBest} (M=8).
  Colours are
  red $(z_1)$,
  blue $(z_2)$,
  green $(z_3)$ and 
  yellow $(z_4)$ \citep[][Eq. 17]{Jet20}.
  Open diamonds denote best frequencies.}
\label{hjd14R412sz}
\end{figure*}

\begin{figure*}
\begin{center}
  \resizebox{12.9cm}{!}
  {\includegraphics{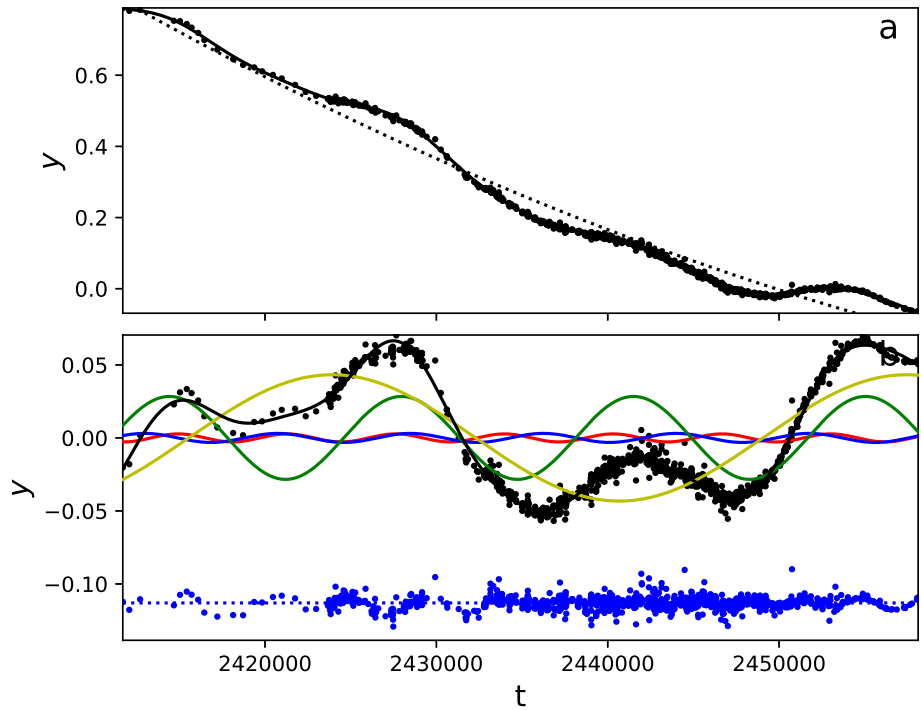}}
\end{center}
\caption{Data and model
  M=8 of Table \ref{TableBest}.
  (a) Data (black dots) and $p(t)$ trend (dotted black line).
  (b) Data minus $p(t)$ trend (black dots),
  $g(t)$ minus $p(t)$ (black line),
  $g_1(t)$ signal (red line),
  $g_2(t)$ signal (blue line),
  $g_3(t)$ signal (green line)
  and
  $g_4(t)$ signal (yellow line).
  Residuals (blue dots) are offset
  to -0.012 (dotted blue line).}
\label{hjd14R412sgdet}
\end{figure*}

\begin{figure*}
\begin{center}
  \resizebox{14.9cm}{!}
  {\includegraphics{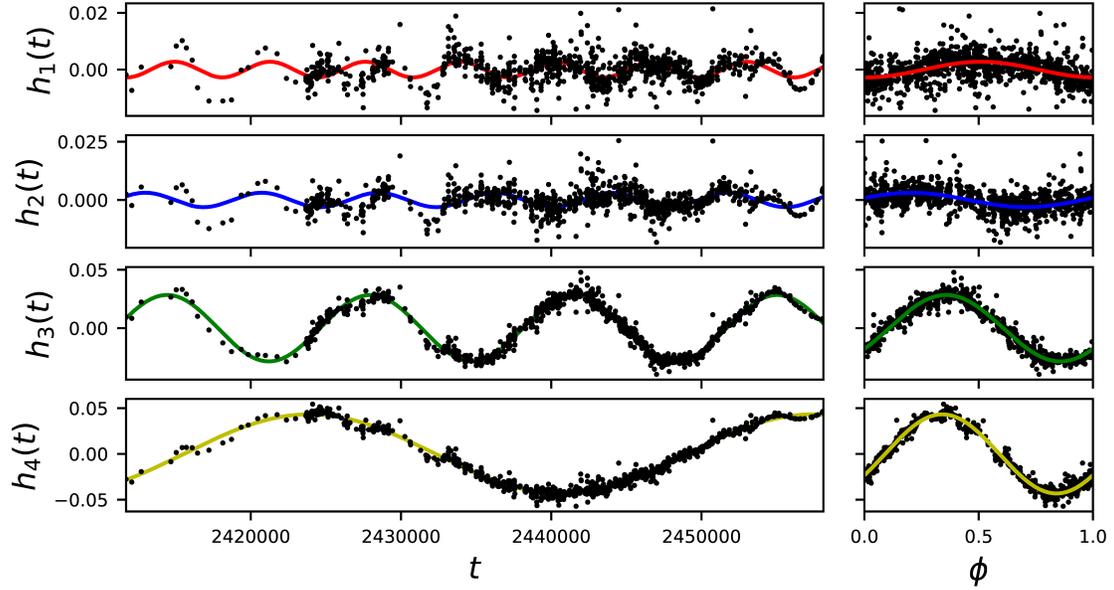}}
\end{center}
\caption{Four signals in original data.
  Signals $y_{i,j}$ (Eq. \ref{EqSignals})
  for model M=8 of
  Table \ref{TableBest}.
  Each signal is plotted as a function
 of time $(t)$ and phase $(\phi)$.}
\label{hjd14R412sSignals}
\end{figure*}

\begin{figure*}
\begin{center}
  \resizebox{12.9cm}{!}
  {\includegraphics{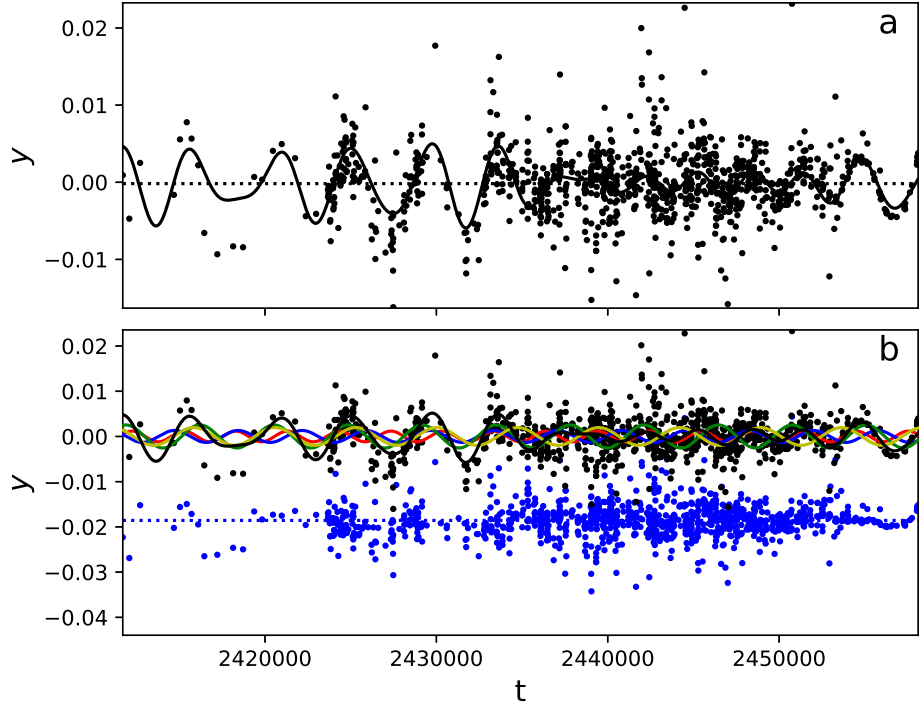}}
\end{center}
\caption{Four signal
  residuals (Table \ref{Tablehjd14}: M=8),
  otherwise as in Fig. \ref{hjd14R412sgdet}.}
\label{hjd58R410sgdet}
\end{figure*}

\begin{figure*}
\begin{center}
  \resizebox{14.9cm}{!}
  {\includegraphics{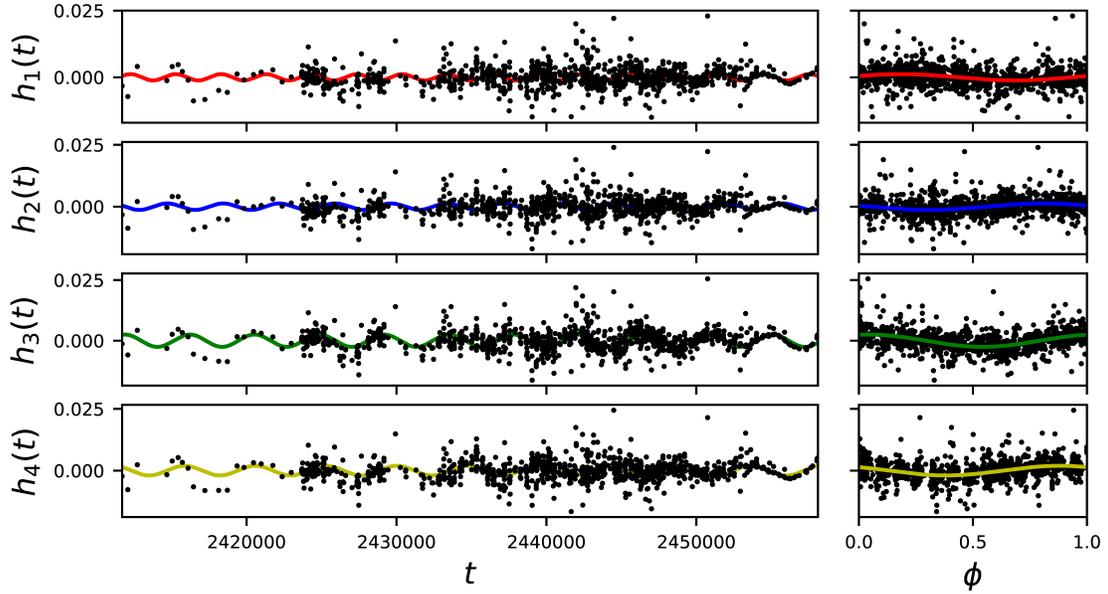}}
\end{center}
\caption{Four signals in four signal residuals.
  Signals $y_{i,j}$ (Eq. \ref{EqSignals})
  for model M=8 in
  Table \ref{Tablehjd14},
  otherwise as in Fig. \ref{hjd14R412sSignals}.}
\label{hjd58R410sSignals}
\end{figure*}

\begin{figure*}
\begin{center}
  \resizebox{12.9cm}{!}
  {\includegraphics{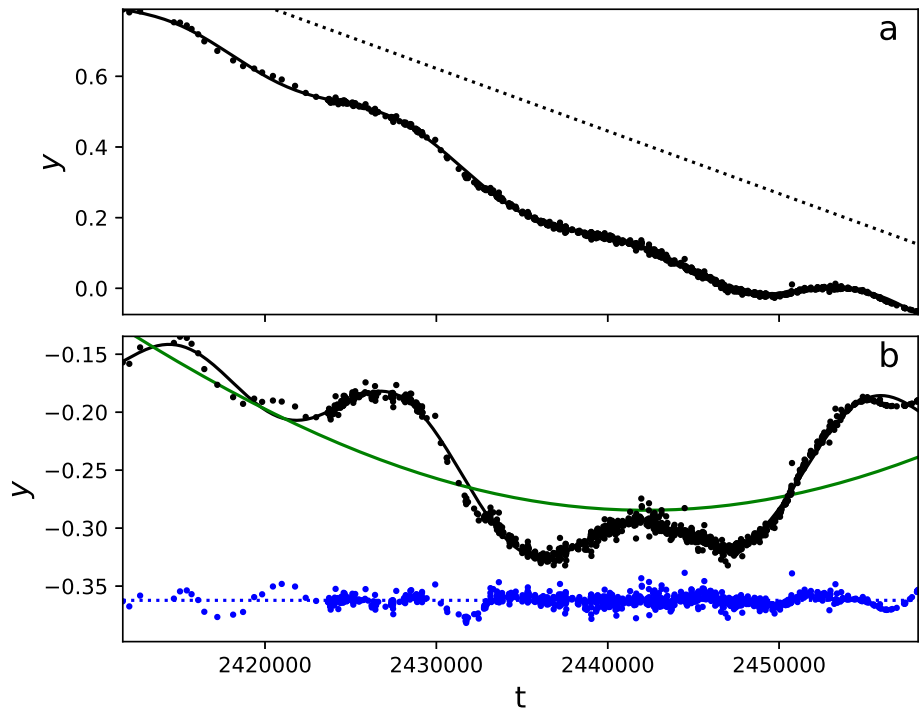}}
\end{center}
\caption{Failed three signal
  model for original data (Table \ref{Tablehjd14}: M=5),
  otherwise as in Fig. \ref{hjd14R412sgdet}.}
\label{hjd14R311sgdet}
\end{figure*}

\begin{figure*}
\begin{center}
  \resizebox{14.9cm}{!}
  {\includegraphics{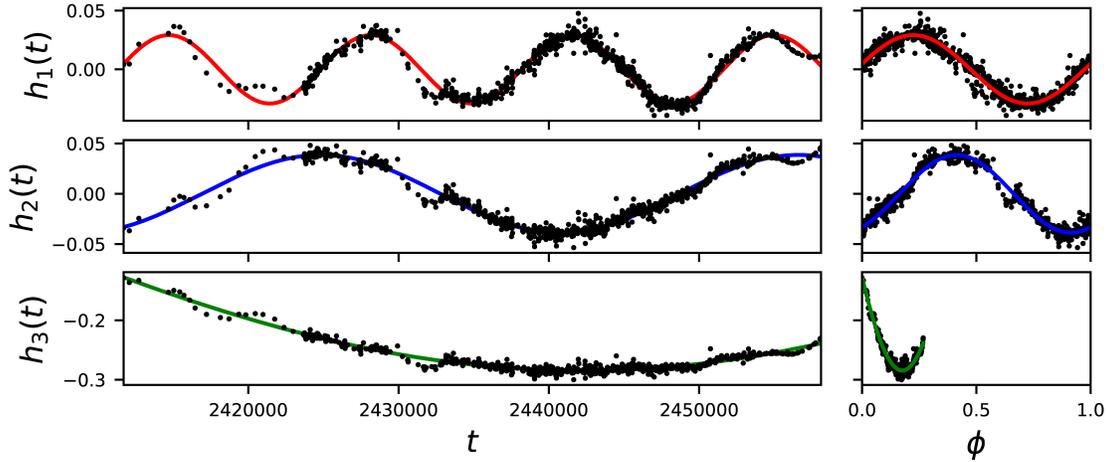}}
\end{center}
\caption{
  Failed model for original data.
  Signals $y_{i,j}$ (Eq. \ref{EqSignals})
  for model M=5 in
  Table \ref{TableBest}. Otherwise
  as in Fig. \ref{hjd14R412sSignals}.}
\label{hjd14R311sSignals}
\end{figure*}

\begin{figure*}
\begin{center}
  \resizebox{14.9cm}{!}
  {\includegraphics{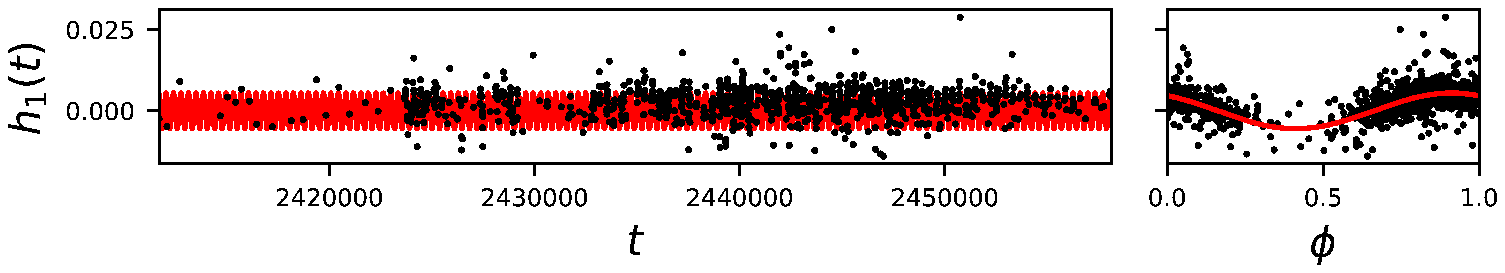}}
\end{center}
\caption{\Ysignal ~in \JDdata.
  Signal $y_{i,j}$ (Eq. \ref{EqSignals})
  for model M=9 in
  Table \ref{TableShort},
  otherwise as in Fig. \ref{jd14R110lSignals}.}
\label{jd14R110lSignals}
\end{figure*}

We analyse the original \HJDdata  ~first, 
because the \Ysignal ~does not contaminate these data.
The one signal model periods $P_1=58237^{\mathrm{d}}$
and $13148^{\mathrm{d}}$
are different (Table \ref{TableBest}: Models M=1 and 2).
Fisher-test reveals that the latter
quadratic $K_3=2$ trend \RModel{1,1,2}
is certainly a better model for the data
than the linear $K_3=1$ trend  \RModel{1,1,1}
(Table \ref{TableBest}: M=1, $\uparrow,F=149, Q_F<10^{-16}$).
Both two signal models M=3 and 4 give
the period $P_1 \approx 13300^{\mathrm{d}}$,
but the $P_2$ periods of these models
are different.
Also for these two signal models, the
$K_3=2$ quadratic trend \RModel{2,1,2} is better than
the linear $K_3=1$ trend \RModel{2,1,1}
(Table \ref{TableBest}: M=3, $\uparrow, F=390, Q_F<10^{-16}$).
The same $\sim 13300^{\mathrm{d}}$ period is also
present in the three signal \RModel{3,1,1} and \RModel{3,1,2},
but the other two periods of these models differ
(Table \ref{TableBest}: M=5 and 6).
Model M=5 fails $(\RD)$, because the two largest
periods have dispersing amplitudes $(\AD)$
and the largest one is unrealistic $(\HD)$.
This linear trend three signal
model M=5 has $p=11$ free parameters.
Yet, the simple quadratic trend two signal model
M=4, having only $p=8$ free parameters,
is certainly a better model for the data
(Table \ref{TableBest}: M=4,
  $\leftarrow, F=0, Q_F=1.0$).
This is very strong evidence for
the $K_3=2$ quadratic $p(t)$ trend in these data.
The linear trend four signal model M=7 fails $(\RD)$.
All $\uparrow$ arrows 
point towards model M=8
in the second last column of
Table \ref{TableBest}.
This four signal \RModel{4,1,2} is certainly
the best one of all eight compared models,
because it beats the other seven
models by an absolute certainty of $Q_F<10^{-16}$.
The periodograms for this \RModel{4,1,2},
and the model itself,
are shown in
Figs. \ref{hjd14R412sz} and
\ref{hjd14R412sgdet}.
The transparent
diamonds denoting the
red $z_1$ and the blue $z_2$ periodogram minima 
for the two weaker periodicities
are certainly real.
When all four periodograms are plotted
in the same scale,
these two minima
appear to be
shallow only because the two
larger amplitude periodic signals clearly
dominate in this model M=8.
These two high amplitude signals have a much bigger
impact on the squared sum of residuals $R$
than the two low amplitude signals.
When the tested frequencies approach zero,
the yellow $z_4$ periodogram turns upwards
in the lower panel of Fig. \ref{hjd14R412sz}.
Thus, DCM can confirm that none of the periods
longer than $\Delta T$ fit to these data.
The level of residuals is stable, but
some regular patterns
indicate
that there are more than
four signals in these data
(Fig. \ref{hjd14R412sgdet}: blue dots).
Each $h_j(t_i)$ signal
\begin{eqnarray}
y_{i,j}=y_i-[g(t_i)-h_j(t_i)]
\label{EqSignals}
\end{eqnarray}
is also shown in Fig. \ref{hjd14R412sSignals}.
The main results in Table \ref{TableBest} are

\begin{itemize}

\item[\ResultOne.] These data contain a
  quadratic $p(t)$ trend $K_3=2$. 

\item[\ResultTwo.] The periods and amplitudes of
  all quadratic trend $K_3=2$ models are consistent
  (Table \ref{TableBest}: M=2, 4, 6 and 8)). When we detect a new signal,
  we re-detect exactly the same
  old signal periods and signal amplitudes.
 
\item[\ResultThree.] There are at least four
  signals, because \RModel{4,1,2} beats the other
  models with an absolute certainty of \REJ.
  
\end{itemize}

\noindent
We use the above 
\ResultOne, \ResultTwo ~and \ResultThree ~abbreviations
for these particular results.
The quadratic $K_3=2$ trend is hereafter used
in all models for the original data
(\ResultOne ~result).

It takes several days
for an ordinary PC
to compute the four signal
\RModel{4,1,2} and
its 20 bootstrap rounds
(Figs.
\ref{hjd14R412sz},
\ref{hjd14R412sgdet}
and
\ref{hjd14R412sSignals}).
The 
five signal model
computation
would take months.
We solve this problem with
\threetest ~and \fourtest ~approach.
First, the original data
are analysed with the quadratic trend model.
Then, we analyse
the three or four signal residuals
with the constant trend model
(e.g. Fig. \ref{hjd58R410sgdet}).
This gives us the six or eight signal
residuals, which are analysed with
the constant trend model.
We end this process at
$3+3+3+3$
or
$4+4+4=12$ signals.
The results for our three samples are given in
Tables \ref{Tablehjd13} and \ref{Tablehjd14} (\HJDdata),
Tables \ref{Tabled13} and \ref{Tabled14} (\Ddata), and
Tables \ref{Tablejd13} and \ref{Tablejd14} (\JDdata).
All these
results are compared in Table \ref{TableComparison}.

\threetest ~results for
\HJDdata, \Ddata ~and \JDdata ~agree
in Table \ref{TableComparison}.
The same applies to \fourtest ~results.
In all the above mentioned tables, we use the symbol

 \begin{itemize}

  \item[] $\WD$ = Weakest of first nine detected signals

  \end{itemize}
  to denote the periodicity that has the lowest
  amplitude of all nine first detected signals.
  We choose these $\WD$ periods from models
  M=9 (\threetest) and M=12 (\fourtest).
  If these $\WD$ periods were ignored
  in Table \ref{TableComparison},
  the results would be the same for the eight best 
  \threetest ~and \fourtest ~periods.
  The main results in Table \ref{TableComparison} are

  \begin{itemize}

  \item[\ResultFour] These data contain at least
    seven or eight signals.

  \item[\ResultFive] \threetest ~and
    \fourtest ~give the same results
    for the eight best periods in all
    three samples.

  \item[\ResultSix] There is no \Ywindow ~and
    no \Ysignal ~bias,
    because all three samples give the same results.
    
  \end{itemize}

  \noindent
  We use the words   ``at least'' in 
  \ResultFour ~result,
  because \HJDdata ~or \Ddata ~may
  contain even more periodicities.
  For example, 
  Fisher-test does not require
  the rejection of the ninth signal
  in \HJDdata
  ~(Tables \ref{Tablehjd13} and \ref{Tablehjd14}: M=9).
  Another uncertainty is connected to the failed $(\RD)$
  model M=6 in
  Tables \ref{Tablehjd13},
   \ref{Tablehjd14},
   \ref{Tabled13} and
   \ref{Tabled14}.
  The two periods above and below $4500^{\mathrm{d}}$
  show amplitude dispersion $(\AD)$ and
  intersecting frequencies $(\HD)$.
  DCM does not tend to find too many signals,
  but it may find too few \citep[see][Sect. 4.4]{Jet20}.
  However, we can not 
  confirm, if these two
  periods above and below  $4500^{\mathrm{d}}$
  represent two separate signals.

  The results for the four signal residuals are shown
  in Figs. \ref{hjd58R410sgdet} and \ref{hjd58R410sSignals}.
  These two figures ``speak for themselves'',
  because the scatter of the new more accurate
  data is extremely small.
  It seems as if these 
  black dots of new data were glued on the
  $h_i(t)$ curves.
  The scatter
  definitely increases for the older less accurate data.
  The eight signal residuals (blue dots) still show some
  regularity, but only for the more accurate new data.
  One failed model is also shown in Figs. \ref{hjd14R311sgdet}
  and \ref{hjd14R311sSignals}.
  Note that the solution for $p(t)$ (black dotted line)
  makes no sense,
  and nor does the solution for $h_3(t)$
  (green continuous line).

\section{Search for short periods}
\label{SectShort}

The eight best \fourtest ~periods are the
same for \HJDdata, \Ddata ~and
\JDdata ~(R5 result).
Therefore, we search for shorter periods
in the residuals of the
eight signal models M=8 of these three samples.
The tested DCM period
range is between $P_{\mathrm{min}}=300^{\mathrm{d}}$
and  $P_{\mathrm{max}}=3000^{\mathrm{d}}$.
Shorter  $P_{\mathrm{min}}$ values would
only lead to the ``detection'' of spurious
signals, like 1/2, 2/3, 3/4 $\times$
\Ywindow ~or \Ysignal.
The short period search results are given
in Table \ref{TableShort}.

The first period
$P_1=365.^{\mathrm{d}}6\pm 78^{\mathrm{d}}$
in \HJDdata
~comes from the \Ysignal ~period,
although this signal should have been removed from
these Heliocentric Julian Days.
While the \Ywindow ~may cause  this periodicity,
it may also be present,
because the units of some epochs are Julian Days,
i.e. the $\delta t_i$ correction of Eq. \ref{EqOCJD}
has not been applied.
The amplitude 
$A_2=0.^{\mathrm{d}}0013$
of the latter second $P_2=616.^{\mathrm{d}}6$ signal
is approximately equal to
the amplitudes of weakest
nine first detected long
period signals  $(\WD)$ in Table \ref{TableComparison}.
Hence, this second signal
may represent a real periodicity.
The third or the fourth short periodicity alternatives
fail
(Table \ref{TableShort}: M=3 and 4 have  ``$\RD$'').

Fisher-test indicates that there may be even
three short period signals
in the eight signal residuals of
\Ddata ~(Table \ref{TableShort}: models M=5-7).
The first one, $P_1=364.^{\mathrm{d}}9\pm 33^{\mathrm{d}}$,
is again the \Ysignal.
This time the \Ywindow ~can not explain this
periodicity, because this window is removed
from \Ddata ~(Eq. \ref{EqTstar}).
The only realistic cause for this periodicity
is the wrong Julian Day units of some epochs.
There can not be only a few such wrong epochs,
because the $\delta t_i$ random shifts of Eq. \ref{EqTstar}
would otherwise eliminate this \Ysignal.
The second $P_2=593.^{\mathrm{d}}1\pm 26^{\mathrm{d}}$ period
is equal to the $P_2$ period already detected from
\HJDdata. It also has the
same amplitude $A_2=0.^{\mathrm{d}}0013$.
The third $P_3=2244^{\mathrm{d}}\pm 19^{\mathrm{d}}$
signal is even stronger, $A_3=0.^{\mathrm{d}}0016$.
There is no fourth signal,
because model M=8 fails $(\RD)$.

Only one period dominates the eight signal
residuals of  \JDdata:
\Ysignal ~(Table \ref{TableShort}: other
models M=10-12 have ``$\RD$'').
This regular artificial signal is shown in Fig. \ref{jd14R110lSignals}.
DCM detects an extremely accurate value for this period,
$P_1=365.^{\mathrm{d}}34\pm 0.^{\mathrm{d}}08$.
The amplitude $A_1\!=\!0^{\mathrm{d}}.0110\pm0.^{\mathrm{d}}0004$
of this \Ysignal ~agrees perfectly with the expected
superimposed signal amplitude
value
$\pm 0.^{\mathrm{d}}0051 \! \equiv \! 0.^{\mathrm{d}}0102$
(Eq. \ref{EqOCJD}).
Our unambiguous re-detection of this superimposed
\Ysignal ~is possible {\it only if} all earlier eight
long period signals
detected from \JDdata ~are also real.
Even a single wrong long period $h_i(t)$ signal could
weaken or erase this \Ysignal,
let alone many wrong
long period $h_i(t)$ signals
or a wrong $p(t)$ trend.
The unexpected \Ysignal ~signatures
in \HJDdata ~and \Ddata ~also support
this same detection of eight real long periods,
because some epochs probably have
the wrong units, Julian Days. 
The main results in Table \ref{TableShort} are

\begin{itemize}

\item[\ResultSeven.] The unambiguous
  detection of the artificially
  superimposed \Ysignal ~in
  \JDdata ~strongly supports the idea
  that the eight detected
  long periods of \target ~are real.

\item[\ResultEight.] Signatures of 
  $\sim 600^{\mathrm{d}}$ signal are
  detected in all three samples.
  \Ddata ~$2244^{\mathrm{d}}$ signal
  may also be a real periodicity.
  
\end{itemize}

\section{Discussion}

\subsection{Period analysis}

While it is not possible to determine the {\it exact}
number of stars
in \target, there are definitely many.
For example, the current O-C data can not confirm
if the two signals below and above $4500^{\mathrm{d}}$
represent one or two periodicities/stars
(e.g. Table \ref{Tablehjd13}: M=6).
Another example is the short $2244^{\mathrm{d}}$ period
detected in \Ddata, which is not detected
in \HJDdata ~or \JDdata
~(Table \ref{TableShort}: M=7).
This detection could be explained with \Ywindow,
which has been removed only from the \Ddata.
There are also Fisher-test
cases that resemble the children's game
stone-paper-scissors, where stone wins scissors,
paper wins stone, and scissors win paper
(e.g. Table \ref{TableShort}: models M=1-4 and M=9-12).
The results for these tests also depend on the
chosen pre-assigned critical level $\gamma_F$ in
the criterion of Eq. \ref{EqFisher}.
In all long period searches, the $Q_F$ critical levels
are extremely significant for seven, eight or even
nine first detected signals. Then this significance
drops abruptly
for the next signals. Exactly the same results were
obtained for simulated data by \citet[][their Table 4]{Jet20}.
DCM may find too few signals,
especially if the $p(t)$ trend is wrong.
However, DCM tends not to detect too many signals,
because such models fail $(\RD)$.

DCM analysis does not require access
to super computers.
The results for \threetest ~and \fourtest ~confirm that
the computation capacity of small PCs is sufficient
for detecting many signals in the O-C data of EBs.
The computations for the bootstrap error estimates
of three and four signal models take a long time.
The best alternative is to run only a few bootstrap
rounds first, and test many bootstrap rounds
only if the preliminary analysis results make sense.
Note that the bootstrap never gives exactly the same error
estimates because it uses random samples of residuals.

We do not re-discuss our
main results \ResultOne ~- \ResultEight.
The fact that DCM re-detects the artificial
\Ysignal ~in \JDdata ~proves beyond any
reasonable doubt that the other detected signals
represent real periodicities.
We succeed in this {\it after} removing a quadratic
$p(t)$ trend and eight $h_i(t)$ signals.

The model used by \citet[][their Eq. 11]{Haj19}
was the sum of a parabola and a sinusoid.
Their  detection rate of fourth bodies in
EBs was only 0.00005.
\citet{Jet19a} has already shown
that the one-dimensional
period finding methods, like the power spectrum
method \citep{Lom76,Sca82,Zec09},
give spurious results if the data
contains many signals.
This applies also to the one-dimensional
periodic model applied by \citet{Haj19},
and explains their low detection rate
of fourth bodies.
At the same time,
their low detection rate
merely highlights the
potential of DCM.


\begin{table}
  \caption{Masses $m_3$ from
    $p_3$ and $a$ of Eq. \ref{EqMasses}.
    Eight periods are from long search (Long)
    and two periods from short search (Short).
    We refer to these stars as S$_1$, ..., S$_{10}$.
    Masses are computed for $i=90^{\mathrm{o}}$
    , $60^{\mathrm{o}}$ and $30^{\mathrm{o}}$
($m_3^{i=90}$, $m_3^{i=60}$, $m_3^{i=30}$). 
  }
    \addtolength{\tabcolsep}{-0.08cm}
\begin{center}
\begin{tabular}{llrrcccc}
  \hline
  &         & \multicolumn{2}{c}{$p_3$} &   $a$    &$m_3^{i=90}$&$m_3^{i=60}$&$m_3^{i=30}$ \\
  \cline{3-4} \cline{6-8}
Search & ~~Star    &      [d ]      &  [y]    &    [d]    &$[m_{\odot}]$&$[m_{\odot}]$&$[m_{\odot}]$\\       
  \hline
  Long & S$_{ 1}$ &        33507 &         91.7 &      0.04300 &   1.16 &   1.38 &   2.74 \\
  Long & S$_{ 2}$ &        13537 &         37.1 &      0.02840 &   1.45 &   1.73 &   3.56 \\
  Long & S$_{ 3}$ &         7754 &         21.2 &      0.00310 &   0.20 &   0.23 &   0.40 \\
  Long & S$_{ 4}$ &         6346 &         17.4 &      0.00275 &   0.20 &   0.23 &   0.41 \\
  Long & S$_{ 5}$ &         4723 &         12.9 &      0.00200 &   0.18 &   0.20 &   0.36 \\
  Long & S$_{ 6}$ &         4320 &         11.8 &      0.00250 &   0.24 &   0.27 &   0.49 \\
  Long & S$_{ 7}$ &         3732 &         10.2 &      0.00130 &   0.13 &   0.15 &   0.27 \\
  Long & S$_{ 8}$ &         3019 &          8.3 &      0.00120 &   0.14 &   0.16 &   0.29 \\
 Short & S$_{ 9}$ &         2244 &          6.1 &      0.00080 &   0.11 &   0.13 &   0.23 \\
 Short & S$_{10}$ &          592 &          1.6 &      0.00060 &   0.21 &   0.25 &   0.44 \\
\hline
\end{tabular}
\end{center}
\addtolength{\tabcolsep}{+0.08cm}
\label{TableMasses}
\end{table}

\citet[][Sect. 6.]{Jet20} already discussed the
problems arising from the use of  $K_2=2$ double
wave models. All third bodies do not necessarily
induce purely sinusoidal O-C changes. Therefore, the
temptation for using  $K_2=2$ harmonics arises.
This approach opens up a real Pandora's box,
because two stars having  periods $P_1$ and $P_2$
induce a synodic period $(P_1^{-1}-P_2^{-1})^{-1}$.
These synodic periods are repeated through out
the whole data, and DCM certainly detects them.
If we had used  the $K_2=2$ option in our analysis,
the interactions between the signals of all stars
in \target ~would have caused an incredible mess.

\subsection{Astrophysics}

Activity cycles are quasi-periodic,
and never give regular long-term residuals.
Therefore, the \citet{App92} mechanism can not explain 
the numerous O-C periods of \target.
An apsidal motion can cause only one period.
If the light-time effect (LTE) of a third
body causes these
periodic O-C changes,
the mass function fulfills
\begin{eqnarray}
f(m_3)=
  {
  {(m_3 \sin{i})^3}
  \over
  {[m_1(1+q)+m_3]^2}
  }
  =
  {
  {(713.15 ~a)^3}
  \over
  {p_{3}^2}
  },
\label{EqMasses}
\end{eqnarray}
where
$i$ is the inclination of the orbital plane of the
third body,
$m_1$ is the mass of primary $[m_{\odot}]$,
$q=m_2/m_1$ is the dimensionless mass ratio of secondary and primary, 
$m_3$ is the mass of the third body $[m_{\odot}]$,
$a=A/2$ is half of the peak the peak amplitude of
O-C modulation caused by the third body  $[{\mathrm{d}}]$,
and
$p_3$ is the period of the modulations 
caused by the third body $[{\mathrm{y}}]$
\citep[][]{Bor96,Tan15,Yan16}.
We use the masses $m_1=3.2 m_{\odot}$
and $m_2=1.3 m_{\odot}$ \citep[][A4~IV + G5~IV]{Dem95}.
The long period search 
$p$ and $a$ values are those
detected from \HJDdata ~(Table \ref{Tablehjd14}: M=4 and 8).
The respective two short period values are detected
from \HJDdata ~and \Ddata ~(Table \ref{TableShort}: M=4 and 7).
We compute the $m_3$ values for inclinations
$i=90^{\mathrm{o}}$,
$60^{\mathrm{o}}$ and $30^{\mathrm{o}}$ (Table \ref{TableMasses}).
In the $i=90^{\mathrm{o}}$ alternative,
the mass $1.45 m_{\odot}$ of star S$_2$ exceeds
the $m_2=1.3 m_{\odot}$ mass of the secondary.
The $1.16 m_{\odot}$ mass of S$_1$ star is just below
this limit. If the remaining eight less massive
stars were in the main sequence, they would all belong
to spectral type M.
The $m_1=3.2 m_{\odot}$ primary would dominate
the luminosity of such a system of twelve stars.
If the inclination were $i=60^{\mathrm{o}}$,
the S$_1$ and $S_2$ star masses would exceed
the secondary $m_2$ mass, but the $m_1$
primary would still dominate the luminosity
of the whole system.
The $i=30^{\mathrm{o}}$ alternative can
be ruled out, because the S$_2$ star would be more
massive and brighter than the primary,
and this practically constant short-term
radial velocity star would have
been noticed long ago.
The orbital period $37.^{\mathrm{y}}1$ of this
S$_2$ star has also been detected
in several previous studies (Table \ref{TableSome}).

\begin{table}
\caption{Some earlier O - C cycle detections. }
\begin{center}
\begin{tabular}{rrl}
\hline
  [y]~~ & [d]~~    & Reference\\
\hline
137.5 & 50222 & \citet{Dem95} \\ 
36.8  & 13441 & \citet{Dem95} \\
11.2  &  4091 & \citet{Dem95} \\
34.8  & 12711 & \citet{Man16} \\
23.3  &  8510 & \citet{Man16} \\
38    & 13879 & \citet{Cha19} \\
\hline
\end{tabular}
\end{center}
\label{TableSome}
\end{table}

The quadratic polynomial trend is
\begin{eqnarray}
  p(t) & = & M_0 + M_1 c ~t + M_2 ~c^2 ~t^2,
\nonumber
\end{eqnarray}
where $c=2/\Delta T$.
The time derivative is
\begin{eqnarray}
  {
  {\mathrm{d} p(t)}
  \over
  {\mathrm{d} t}
  }
  = M_1 c +  2 M_2c^2 ~t.
 \nonumber
\end{eqnarray}
\noindent
We use the values 
$M_1=-0.627 \pm 0.005$
and
$M_2=0.082\pm 0.002$
of the four signal
model for the original
\HJDdata ~(Table \ref{TableM}: M=4).
The $p(t)$ changes caused by  parameter $M_1 c$
can be eliminated by computing the O-C values
with a new constant
period $P'=P - (M_1 c)/P=1.^{\mathrm{d}}3573458$.
Hence, the only real period
change is $\Delta P = 2 M_2 c^2  =
4.^{\mathrm{d}}1 \times 10^{-10}=3.^{\mathrm{s}}6 \times 10^{-5}$ 
during $P=1.^{\mathrm{d}}335730911$.
The period increases because $M_2>0$.
The long-term increase rate is
\begin{eqnarray}
{{\Delta P} \over { P} }=2 M_2 c^2 = (3.04 \pm 0.07) \times10^{-10},
\end{eqnarray}
which is about
five times less than the $\Delta P/P=1.45\times10^{-9}$
estimate of \target ~obtained
by \citet{Man16}. 
The two and the three signal model results would
have been nearly the same, but the one signal model
$M_1$ and $M_2$ values
would have given a completely wrong result
(Table \ref{TableM}).
We do not derive any mass transfer estimate for \target
~\citep[e.g.][their Eq. 8]{Man16}.
This estimate would not be correct for
a multiple star system, where the
numerous WOSs can perturb
the central EB, e.g.  through
the  Kozai effect \citep{Koz62},
or the combination of Kozai cycle and
tidal friction effects \citep{Fab07}.

Our solar system is stable because the orbital
planes of planets are nearly co-planar, the
biggest exception being the smallest planet
Mercury with an inclination of seven degrees.
The single stars, the multiple star systems
and the planets 
form in co-planar
protostellar disks \citep[e.g.][]{Wat98}.
If the orbital plane of a third body
is not co-planar with, or perpendicular to,
the orbital plane of the central EB,
periodic long-term perturbations change the
orbital plane of the central EB,
and the eclipses may no longer occur
\citep[][Eq. 27]{Sod75}.
Since no such effects have been observed in \target ~in
over a century, the planes of {\it all} WOSs
are most probably co-planar.
There orbital plane of central EB is stable for
$\Psi=0^{\mathrm{o}}$ or $90^{\mathrm{o}}$,
where $\Psi$ is the angle between central EB orbital plane
and WOSs orbital plane.
This is the case for Algol
\citep[][Algol~AB and Algol~C have
$\Psi=90.^{\mathrm{o}}20\pm0.^{\mathrm{o}}32$]{Bar12},
where 
two new wide orbit stars Algol~D and Algol~E
were recently detected by \citet{Jet20B}.

\section{Conclusions}

\citet{Haj19} detected only four candidates possibly
having a fourth body in their O-C data of 80~000
eclipsing binaries.
The probability for
detecting a fourth body from their O-C data was
only 0.00005.
Here, we apply the new Discrete Chi-Square Method (DCM)
to the O-C data of the eclipsing binary \target,
and detect signatures of at least ten wide orbit
stars (WOSs) orbiting
the central EB.
These WOSs have orbital periods
between $1.^{\mathrm{y}}6$
and $91.^{\mathrm{y}}7$.
Two WOSs are certainly
more massive
than the Sun.
The orbits of {\it all} these WOSs
are most probably co-planar with, or perpendicular to,
the orbital plane
of central EB, because no
changes have been observed in the
eclipses of \target ~in over a century.
Considering the number of new companions
detected in \target, it is actually a more interesting
multiple star system
than Algol itself \citep[][``only'' two
new companions
Algol D and Algol E]{Jet20B}.
Our results for \target ~and Algol confirm
that many EBs have unknown companions,
which can be easily detected with our DCM.
This abstract mathematical method is not
designed for analysing any particular
phenomena of the real physical world.
The O-C data of EBs just happen to be
one particular type of suitable data.
We sincerely hope that
these results
would ignite a ``renaissance'' in
the O-C data studies of other EBs.
Valuable data have been patiently
collected by the professional and
the amateur astronomers since the well-known
amateur astronomer Sir
John Goodricke re-detected 
Algol's periodic eclipses 
\citep{Goo83}.

~ \\ ~ \\
{\bf Acknowlegdements.}
This work has made use of NASA's 
Astrophysics Data System (ADS) services.
We retrieved the
O-C data of \target ~from the
Lichtenknecker Database of the BAV.
\bibliographystyle{apalike}


\newcommand{\Pmuutos}{
$p(t)=M_0 [2t/\Delta T]^0  +M_1 [2t/\Delta T]^1+M_2[2t/\Delta T]^2$ \\
$    = M_0 + M_1 [2/\Delta T] t + M_2 [2/\Delta T]^2 t^2$ \\
$    = M_0 + M_1 c ~t + M_2 c^2 ~t^2 $, where $c=2/\Delta T$ \\
$\rightarrow d p(t) / dt = M_1 c  + 2 M_2 c^2 t$\\
$= a + b t$, where $a=M_1 c$ and $b=2 M_2 c^2$ \\
$M_0 =8.09261e-01 \pm 2.41590e-03$ \\
$M_1=-6.26763e-01 \pm 4.87297e-03$ \\ 
$M_2=  8.15231e-02 \pm 1.98385e-03$ \\
$\Delta T = 4.6410973700e+04$ \\
$\partial (2 M_2 c^2)/\partial M_2= 2 c^2$
$\rightarrow \sigma_{\Delta P/P}=\sqrt{ 4 c^4 \sigma_{M_2}^2}$
}

\newcommand{\DOC}{
    \noindent
  $a=A/2$ \\
  $\sigma_a=\sigma_A/2$ \\
  $p=P/2$ \\
  $\sigma_p=\sigma_P/2$\\
$C=f(m)= (713.15 ~a)^3/p^2$  \\
$D=\partial C/\partial a = 3 (713.15)^3 a^2/p^2$ \\
$E=\partial C/\partial p = -2 (713.15 ~a)^3/p^3$ \\
$\sigma_C= \sqrt{(D \sigma_a)^2 + (E \sigma_p)^2}$}

\appendix
\setcounter{table}{0}
\setcounter{figure}{0}
\renewcommand{\thetable}{A\arabic{table}}
\renewcommand{\thefigure}{A\arabic{figure}}

\section{Reproducing our results}

In this appendix, we give all necessary information for
reproducing the results of our DCM period
analysis. DCM manual \PR{manual.pdf}, 
and all other required analysis files, can be copied
from the Zenodo database.
Copy the analysis files to the same folder
in your computer.
Do not edit these analysis files.

The DCM analysis program file is

\begin{itemize}

  \item[] \PR{dcm.py}

  \end{itemize}
  
The three different \PR{file1} data files are

\begin{itemize}

  \item[] \PR{hjdXZAnd.dat} = \HJDdata ~= Table \ref{TableHJD}
  \item[] \PR{dXZAnd.dat}   = \Ddata ~= Table \ref{TableD}
  \item[] \PR{jdXZAnd.dat}  = \JDdata ~= Table \ref{TableJD}

\end{itemize}

There are {\color{red} 98} control files,
which are specified in the last columns of
Tables
\ref{Tablehjd13},
\ref{Tablehjd14},
\ref{Tabled13},
\ref{Tabled14},
\ref{Tablejd13}, 
\ref{Tablejd14}
and
\ref{TableShort}.
All our results can reproduced by
using these control files.
For example, the two linux
commands
\begin{itemize}
\item[] \PR{cp hjd14R111s.dat dcm.dat}
\item[] \PR{python dcm.py}
\end{itemize}
reproduce \RModel{1,1,1}
period analysis
results given in the first line
of Table \ref{TableBest} (M=1).
In other words, the control file
\PR{dcm.dat} is an exact copy of \PR{hdj14R111s.dat}.
This control file specifies what
kind of a DCM analysis the program \PR{dcm.py}
should perform.
The meaning of all control file \PR{dcm.dat} parameters
is explained in the manual \PR{manual.pdf}.

The naming conventions of our control files
are given in Table \ref{Tabledcm}.
These names are formed by using a sequence N1 + N2 + N3 + N4 + N5.
The first N1 notation \PR{hjd} in the file name
\PR{hjd14R111s.dat} means that we analyse
\HJDdata.
The N2 notation \PR{14} means that we search for
the four first signals in the original data.
The N3 notation \PR{R} indicates that the DCM test
statistic $z$ is computed from the sum
of squared residuals  $R$ \citep[][Eqs. 9 and 11]{Jet20}.
The N4 notation  \PR{111} tells that we use \RModel{1,1,1}.
The last N5 notation \PR{s} refers to the search
for small frequencies
(long period search between 3000 and 100~000 days).
All figures and files produced by \PR{dcm.py} begin
with the tag \PR{hjd14R111s}, e.g. the periodogram
figure \PR{hjd14R111sz.eps} or the result file
\PR{hjd14R111sParams.dat}.

DCM analysis program \PR{dcm.py}
can be applied to the O-C data of any other
star, if the format for the data file \PR{file1}
and the control file \PR{dcm.dat} are the same
as in this paper.

\begin{table}
\caption{\PR{dcm.dat} control file name notations.}
\begin{center}
\addtolength{\tabcolsep}{-0.12cm}
\begin{tabular}{clccccc}
\hline
  & Meaning                              &    N1  & N2     &    N3  & N4       & N5 \\
\hline
\PR{hjd}  & \HJDdata                     &\PR{hjd}& \PR{*} & \PR{*} & \PR{*}   & \PR{*} \\
\PR{jd}   & \Ddata                       & \PR{jd}& \PR{*} & \PR{*} & \PR{*}   & \PR{*} \\ 
\PR{jd}   & \JDdata                      & \PR{jd}& \PR{*} & \PR{*} & \PR{*}   & \PR{*} \\ 
\PR{14}   & Signals 1-4 for data         & \PR{*} & \PR{14}& \PR{*} & \PR{*}   & \PR{*} \\
\PR{58}   & Signals 5-8 for residuals    & \PR{*} & \PR{58}& \PR{*} & \PR{*}   & \PR{*} \\
\PR{912}  & Signals 9-12 for residuals   & \PR{*} &\PR{912}& \PR{*} & \PR{*}   & \PR{*} \\
\PR{13}   & Signals 1-3 for data         & \PR{*} & \PR{13}& \PR{*} & \PR{*}   & \PR{*} \\
\PR{46}   & Signals 4-6 for residuals    & \PR{*} & \PR{46}& \PR{*} & \PR{*}   & \PR{*} \\
\PR{79}   & Signals 7-9 for residuals    & \PR{*} & \PR{79}& \PR{*} & \PR{*}   & \PR{*} \\
\PR{1012} & Signals 10-12 for residuals  & \PR{*}&\PR{1012}& \PR{*} & \PR{*}   & \PR{*} \\
\PR{R}    & $R  $ test statistic         & \PR{*} & \PR{*} & \PR{R} & \PR{*}   & \PR{*} \\
\PR{111}  & \RModel{1,1,1} for data      & \PR{*} & \PR{*} & \PR{*} & \PR{111} & \PR{*} \\
\PR{112}  & \RModel{1,1,2} for data      & \PR{*} & \PR{*} & \PR{*} & \PR{112} & \PR{*} \\
\PR{211}  & \RModel{2,1,1} for data      & \PR{*} & \PR{*} & \PR{*} & \PR{211} & \PR{*} \\
\PR{212}  & \RModel{2,1,2} for data      & \PR{*} & \PR{*} & \PR{*} & \PR{212} & \PR{*} \\
\PR{311}  & \RModel{3,1,1} for data      & \PR{*} & \PR{*} & \PR{*} & \PR{311} & \PR{*} \\
\PR{312}  & \RModel{3,1,2} for data      & \PR{*} & \PR{*} & \PR{*} & \PR{312} & \PR{*} \\
\PR{411}  & \RModel{4,1,1} for data      & \PR{*} & \PR{*} & \PR{*} & \PR{411} & \PR{*} \\
\PR{412}  & \RModel{4,1,2} for data      & \PR{*} & \PR{*} & \PR{*} & \PR{412} & \PR{*} \\
\PR{110}  & \RModel{1,1,0} for residuals & \PR{*} & \PR{*} & \PR{*} & \PR{110} & \PR{*} \\
\PR{210}  & \RModel{2,1,0} for residuals & \PR{*} & \PR{*} & \PR{*} & \PR{210} & \PR{*} \\
\PR{310}  & \RModel{3,1,0} for residuals & \PR{*} & \PR{*} & \PR{*} & \PR{310} & \PR{*} \\
\PR{410}  & \RModel{4,1,0} for residuals & \PR{*} & \PR{*} & \PR{*} & \PR{410} & \PR{*} \\
\PR{s}    & Small frequencies            & \PR{*} & \PR{*} & \PR{*} & \PR{*} & \PR{s} \\
\PR{l}    & Large frequencies            & \PR{*} & \PR{*} & \PR{*} & \PR{*} & \PR{l} \\
\hline
\end{tabular}                                                           
\addtolength{\tabcolsep}{+0.12cm}
\end{center}
\label{Tabledcm}
\end{table}

\begin{table}
  \caption{Rejected data. Date, UT
    and rejection criterion}
\begin{center}
\begin{tabular}{ccl}
\hline
 Date & UT & Criterion \\
\hline
           06.10.1923 &  21:36  & Secondary minimum \\ 
           11.11.1949 &  05:47  & Secondary minimum \\ 
           17.09.1965 &  01:13  & Secondary minimum \\ 
           25.11.1975 &  17:56  & Secondary minimum \\ 
           26.12.1996 &  00:04  & Secondary minimum \\ 
           03.01.2008 &  18:51  & Secondary minimum \\ 
           21.12.2009 &  19:03  & Secondary minimum \\ 
           14.08.2010 &  23:02  & Secondary minimum \\ 
           18.09.1923 &  17:02         & Outlier \\ 
           01.10.1923 &  18:57         & Outlier \\ 
           02.12.1923 &  19:00         & Outlier \\ 
\hline
\end{tabular}
\end{center}
\label{TableRejected}
\end{table}

\begin{table}
  \caption{
    O-C data in file \PR{hjdXZAnd.dat}.
    Columns are primary eclipse epoch in the Sun
    $(t_{\odot})$ and
    observed minus computed eclipse epoch of Eq. \ref{EqOCHJD}
    $(y_{\mathrm{HJD}} \pm \sigma_{y_{\mathrm{HJD}}})$.
  Only three first values
    of all $n=1091$ values are shown.}
\begin{center}
\begin{tabular}{ccc}
\hline
  $t_{\odot}$       &  $y_{\mathrm{HJD}}$   &  $\sigma_{y_{\mathrm{HJD}}}$ \\
  ${\mathrm{[HJD]}}$&  ${\mathrm{[d]}}$     &  ${\mathrm{[d]}}$ \\ 
\hline
       2411700.59000 &   0.789000           &  0.000173            \\
       2412080.62900 &   0.781000           & 0.000173             \\
       2412711.78000 &   0.784000           & 0.000173             \\
       ...           &              ...     &              ...     \\
\hline
\end{tabular}
\end{center}
\label{TableHJD}
\end{table}

\begin{table}
\caption{Yearly distribution of data}
\addtolength{\tabcolsep}{-0.12cm}
\begin{center}
\begin{tabular}{cccccccccccc}
\hline
Jan &
Feb &
Mar &
Apr &
May &
Jun &
Jul &
Aug &
Sep &
Oct &
Nov &
Dec \\
98 &
62 &
25 &
5 &
4 &
11 &
56 &
142 &
220 &
197 &
153 &
121 \\
\hline
\end{tabular}
\end{center}
\addtolength{\tabcolsep}{+0.12cm}
\label{TableMonths}
\end{table}

\begin{table}
  \caption{O-C data in file \PR{dXZAnd.dat}.
    Time points computed from Eq. \ref{EqTstar},
    otherwise as in Table \ref{TableHJD}.}
\begin{center}
\begin{tabular}{ccc}
\hline
       2411655.32320&     0.789000  &     0.000173 \\
       2411965.26421&     0.781000  &     0.000173 \\
       2412702.44481&     0.784000  &     0.000173 \\
      ...           &     ...       &     ...     \\
\hline
\end{tabular}
\end{center}
\label{TableD}
\end{table}

\begin{table}
  \caption{O-C data in file \PR{jdXZAnd.dat}.
    Columns are primary eclipse epoch on the Earth
    $(t_{\oplus})$ and
    observed minus computed eclipse epoch of Eq. \ref{EqOCJD}
    $(y_{\mathrm{JD}} \pm \sigma_{y_{\mathrm{JD}}})$,
    otherwise as in Table \ref{TableHJD}.   }
\begin{center}
\begin{tabular}{ccc}
  \hline
  $t_{\oplus}$      &  $y_{\mathrm{JD}}$   &  $\sigma_{y_{\mathrm{JD}}}$ \\
  ${\mathrm{[JD]}}$ &  ${\mathrm{[d]}}$    &  ${\mathrm{[d]}}$ \\ 
\hline
      2411700.590000 &   0.793510    &    0.000173 \\
      2412080.629000 &   0.784770    &    0.000173 \\
      2412711.780000 &   0.786750    &    0.000173 \\
       ...           &              ...     &              ...     \\
\hline
\end{tabular}
\end{center}
\label{TableJD}
\end{table}

\begin{table*}
  \caption{Long period search between $3000^{\mathrm{d}}$ and $100000^{\mathrm{d}}$
    for \HJDdata.
    Col 1.   Model number M.
    Col 2.   \RModel{K_1,K_2,K_3}, $p=$ number of free parameters and $R=$ sum of squared residuals.
    Cols 3-6.  Period analysis results: Detected periods $P_1, ..., P_4$
    and amplitudes $A_1, ..., A_4$.
    Cols 7-13. Fisher test results:
    ``$\uparrow$ $" \equiv$ complex model above is better than left side simple model,
    ``$\leftarrow$" $\equiv$ left side simple model is better than complex model above,
    $F=$ Fisher test statistic
    and 
    $Q_F=$ critical level.
    Col 14. Figure numbers and control file \PR{dcm.dat}.
    Notations
    ``$\AD$'',
    ``$\HD$'' and
    ``$\RD$''
      are 
      explained in Sect. \ref{SectMethod}.  }
\begin{tiny}
\begin{center}
\addtolength{\tabcolsep}{-0.21cm}
\begin{tabular}{cccccccccccccccc}
  \hline
 1.& 2.          & 3.                 & 4.               & 5.              & 6.             && 7.           & 8.           & 9.           & 10.           & 11.           & 12.           & 13.          & 14.\\
  \multicolumn{15}{c}{Original data (Table \ref{TableHJD} = \PR{file1} = \PR{hjdXZAnd.dat})} \\
 &               & \multicolumn{4}{c}{Period analysis}                                    &~~~~& \multicolumn{7}{c}{Fisher-test}                                                                         &                                                  \\
                 \cline{3-6}                                                                 \cline{8-14}
M& Model          & $P_1 \& A_1 \Dd   $&$P_2 \& A_2  \Dd $&$P_3 \& A_3 \Dd $&$P_4 \& A_4\Dd $&&\RModel{1,1,2}&\RModel{2,1,1}&\RModel{2,1,2}&\RModel{3,1,1}&\RModel{3,1,2} &\RModel{4,1,1} &\RModel{4,1,2}&                                                  \\
\hline
1&\RModel{1,1,1} &$58237\PM2998     $&     -            & -               & -              &&$\uparrow     $&$\uparrow    $&$\uparrow    $&$\uparrow     $&$\uparrow     $&$\uparrow    $&$\uparrow     $& \\ 
 &$p\ON5        $&$0.24 \PM0.02     $&     -            & -                & -              &&$F\ON149     $&$F\ON4260    $&$F\ON4436    $&$F\ON2952     $&$F\ON2855     $&$F\ON2222     $&$F\ON2296    $& \PR{dcm.dat} = \PR{hjd14R111s.dat}               \\
 &$R  \ON0.4089$ &                   &                  &                 &                &&\REJ           &\REJ          &\REJ          &\REJ           &\REJ           &\REJ           &\REJ          &                                                  \\  
  \hline
2&\RModel{1,1,2} &$13148\PM 109      $&     -            & -               & -              && -            &$\uparrow    $&$\uparrow    $&$\uparrow     $&$\uparrow     $&$\uparrow     $&$\uparrow    $& \\
\hline
\end{tabular}
\addtolength{\tabcolsep}{+0.21cm}
\label{TableBest}
\end{center}
\end{tiny}
\end{table*}

\begin{table*}
  \caption{Long period search between $3000^{\mathrm{d}}$ and $100000^{\mathrm{d}}$.
    \threetest ~for \HJDdata. Notation ``$\WD$'' is explained
    in Sect \ref{SectLong}, otherwise as in Table \ref{TableBest}.}
\begin{tiny}
\begin{center}
\addtolength{\tabcolsep}{-0.18cm}
\begin{tabular}{clcccccccc}
\hline
 \multicolumn{9}{c}{Original data (Table \ref{TableHJD} = \PR{file1} = \PR{hjdXZAnd.dat}))} \\
 &              & \multicolumn{3}{c}{Period analysis}                     &~~~~& \multicolumn{2}{c}{Fisher-test} &                                            \\
                 \cline{3-5}                                                                 \cline{7-8}
M&Model          & $P_1 \& A_1  \Dd  $&$P_2 \& A_2   \Dd$&$P_3 \& A_3 \Dd $&&\RModel{2,1,2}&\RModel{3,1,2} &                                                  \\
\hline
1&\RModel{1,1,2} &$13148\PM  92      $&     -            & -               &&$\uparrow    $&$\uparrow     $& \\ 
 &$p\ON6        $&$0.050\PM0.002     $&     -            & -               &&$F\ON5157    $&$F\ON2907     $& \PR{dcm.dat} = \PR{hjd13R112s.dat}               \\
 &$R  \ON0.3595$ &                    &                  &                 &&\REJ          &\REJ           &                                                  \\ 
\hline
 2&\RModel{2,1,2} &$13420\PM 21       $&$32196\PM 335    $& -               && -            &$\uparrow     $& \\ 
 &$p\ON9        $&$0.0582\PM0.0005   $&$0.081\PM0.002   $& -               && -            &$F\ON44.0     $& \PR{dcm.dat} = \PR{hjd13R212s.dat}               \\
\end{tabular}
\addtolength{\tabcolsep}{+0.18cm}
\label{Tablehjd13}
\end{center}
\end{tiny}
\end{table*}

\begin{table*}
  \caption{Long period search between $3000^{\mathrm{d}}$ and $100000^{\mathrm{d}}$.
    \fourtest ~for \HJDdata, otherwise as in Table \ref{TableBest}.} 
\begin{tiny}
\begin{center}
\addtolength{\tabcolsep}{-0.18cm}
\begin{tabular}{clccccccccccc}
\hline
\multicolumn{11}{c}{Original data (Table \ref{TableHJD} = \PR{file1} = \PR{hjdXZAnd.dat})} \\
 &               & \multicolumn{4}{c}{Period analysis}                                    &~~~~& \multicolumn{3}{c}{Fisher-test}           &                                                  \\
                 \cline{3-6}                                                                 \cline{8-10}
M& Model         & $P_1 \& A_1    \Dd$&$P_2 \& A_2   \Dd$&$P_3 \& A_3  \Dd$&$P_4 \& A_4 \Dd$&&\RModel{2,1,2}&\RModel{3,1,2}  &\RModel{4,1,2}&                                                  \\
\hline
1&\RModel{1,1,2} &$13148\PM 109      $&     -            & -               & -              &&$\uparrow    $&$\uparrow      $&$\uparrow    $& \\ 
 &$p\ON6        $&$0.050 \PM0.001    $&     -            & -               & -              &&$F\ON5157    $&$F\ON2907      $&$F\ON2228    $& \PR{dcm.dat} = \PR{hjd14R112s.dat}               \\
 &$R  \ON0.3595$ &                    &                  &                 &                &&\REJ          &\REJ            &\REJ          &                                                  \\ 
\hline
2&\RModel{2,1,2} &$13420\PM28        $&$32196\pm 259    $& -               & -              && -            &$\uparrow      $&$\uparrow    $& \\ 
 &$p\ON9        $&$0.0582\PM0.0004   $&$0.081\PM0.001   $& -               & -              && -            &$F\ON44.0      $&$F\ON50.9    $& \PR{dcm.dat} = \PR{hjd14R212s.dat}               \\
\end{tabular}
\addtolength{\tabcolsep}{+0.18cm}
\label{Tablehjd14}
\end{center}
\end{tiny}
\end{table*}

\begin{table*}
  \caption{Long period search between $3000^{\mathrm{d}}$ and $100000^{\mathrm{d}}$.
    \threetest ~for \Ddata, otherwise as in Table \ref{TableBest}}
\begin{tiny}
\begin{center}
\addtolength{\tabcolsep}{-0.18cm}
\begin{tabular}{cclccccccc}
\hline
  \multicolumn{9}{c}{Original data (Table \ref{TableD} $\equiv$ \PR{dXZAnd.dat})} \\
 &             & \multicolumn{3}{c}{Period analysis}                     &~~~~& \multicolumn{2}{c}{Fisher-test} &                                          \\
                 \cline{3-5}                                                                 \cline{7-8}
M&Model          & $P_1 \& A_1 \Dd   $&$P_2 \& A_2 \Dd  $&$P_3 \& A_3\Dd  $&&\RModel{2,1,2}&\RModel{3,1,2} &                                               \\
\hline
1&\RModel{1,1,2} &$13168\PM 76       $&     -            & -               &&$\uparrow    $&$\uparrow     $& \\ 
  &$p\ON6        $&$0.050\PM0.002    $&     -            & -               &&$F\ON4794    $&$F\ON2663     $& \PR{dcm.dat} = \PR{d13R112s.dat}              \\
 &$R  \ON0.3607$ &                    &                  &                 &&\REJ          &\REJ           &                                               \\ 
\hline
2&\RModel{2,1,2} &$13428\PM 20       $&$32390\pm 320    $& -               && -            &$\uparrow     $& \\ 
 &$p\ON9        $&$0.0578\PM0.0003   $&$0.082\PM0.001   $& -               && -            &$F\ON38.1     $& \PR{dcm.dat} = \PR{d13R212s.dat}              \\
\end{tabular}
\addtolength{\tabcolsep}{+0.18cm}
\label{Tabled13}
\end{center}
\end{tiny}
\end{table*}

\begin{table*}
  \caption{Long period search between $3000^{\mathrm{d}}$ and $100000^{\mathrm{d}}$.
    \fourtest ~for \Ddata, otherwise as in Table \ref{TableBest}.} 
\begin{tiny}
\begin{center}
\addtolength{\tabcolsep}{-0.18cm}
\begin{tabular}{clccccccccccc}
\hline
\multicolumn{11}{c}{Original data (Table \ref{TableD} = \PR{file1} = \PR{dXZAnd.dat})} \\
 &               & \multicolumn{4}{c}{Period analysis}                                    &~~~~& \multicolumn{3}{c}{Fisher-test}           &                                                  \\
                 \cline{3-6}                                                                 \cline{8-10}
M& Model         & $P_1 \& A_1    \Dd$&$P_2 \& A_2   \Dd$&$P_3 \& A_3  \Dd$&$P_4 \& A_4 \Dd$&&\RModel{2,1,2}&\RModel{3,1,2}  &\RModel{4,1,2}&                                                  \\
\hline
  1&\RModel{1,1,2} &$13168\PM 86     $&     -            & -               & -              &&$\uparrow    $&$\uparrow      $&$\uparrow    $& \\ 
 &$p\ON6        $&$0.050 \PM0.002    $&     -            & -               & -              &&$F\ON4794    $&$F\ON2663      $&$F\ON2026    $& \PR{dcm.dat} = \PR{d14R112s.dat}                 \\
 &$R  \ON0.3607$ &                    &                  &                 &                &&\REJ          &\REJ            &\REJ          &                                                  \\ 
  \hline
2&\RModel{2,1,2} &$13428\PM28        $&$32390\pm 385    $& -               & -              && -            &$\uparrow      $&$\uparrow    $& \\ 
 &$p\ON9        $&$0.0578\PM0.0006   $&$0.082\PM0.001   $& -               & -              && -            &$F\ON38.1      $&$F\ON46.0    $& \PR{dcm.dat} = PR{d14R212s.dat}                  \\
\end{tabular}
\addtolength{\tabcolsep}{+0.18cm}
\label{Tabled14}
\end{center}
\end{tiny}
\end{table*}

\begin{table*}
  \caption{Long period search between $3000^{\mathrm{d}}$ and $100000^{\mathrm{d}}$.
    \threetest ~for \JDdata, otherwise as in Table \ref{TableBest}.}
\begin{tiny}
\begin{center}
\addtolength{\tabcolsep}{-0.18cm}
\begin{tabular}{cclccccccc}
\hline
  \multicolumn{9}{c}{Original data (Table \ref{TableJD} $\equiv$ \PR{jdXZAnd.dat})} \\
 &             & \multicolumn{3}{c}{Period analysis}                     &~~~~& \multicolumn{2}{c}{Fisher-test} &                             \\
                 \cline{3-5}                                                                 \cline{7-8}
M&Model          & $P_1 \& A_1   \Dd $&$P_2 \& A_2 \Dd  $&$P_3 \& A_3\Dd  $&&\RModel{2,1,2}&\RModel{3,1,2} &                                       \\
\hline
1&\RModel{1,1,2} &$13146\PM 101       $&     -            & -              &&$\uparrow    $&$\uparrow     $& \\ 
  &$p\ON6        $&$0.050\PM0.001    $&     -             & -              &&$F\ON3816    $&$F\ON2111     $& \PR{dcm.dat} = \PR{jd13R112s.dat}              \\
  &$R  \ON0.3605$ &                    &                  &                &&\REJ          &\REJ           &                                                \\
\hline
2&\RModel{2,1,2} &$13421\PM 26       $&$31957\pm 268    $& -               && -            &$\uparrow     $& \\ 
 &$p\ON9        $&$0.0585\PM0.0004   $&$0.079\PM0.001   $& -               && -            &$F\ON36.0     $& \PR{dcm.dat} = \PR{jd13R212s.dat}              \\
\end{tabular}
\addtolength{\tabcolsep}{+0.18cm}
\label{Tablejd13}
\end{center}
\end{tiny}
\end{table*}

\begin{table*}
  \caption{Long period search between $3000^{\mathrm{d}}$ and $100000^{\mathrm{d}}$.
    \fourtest ~for \JDdata, otherwise as in Table \ref{TableBest}.} 
\begin{tiny}
\begin{center}
\addtolength{\tabcolsep}{-0.18cm}
\begin{tabular}{clccccccccccc}
\hline
\multicolumn{11}{c}{Original data (Table \ref{TableD} = \PR{file1} = \PR{jdXZAnd.dat})} \\
 &               & \multicolumn{4}{c}{Period analysis}                                    &~~~~& \multicolumn{3}{c}{Fisher-test}           &                                                    \\
                 \cline{3-6}                                                                 \cline{8-10}
M& Model         & $P_1 \& A_1    \Dd$&$P_2 \& A_2   \Dd$&$P_3 \& A_3  \Dd$&$P_4 \& A_4 \Dd$&&\RModel{2,1,2}&\RModel{3,1,2}  &\RModel{4,1,2}&                                                   \\
\hline
1&\RModel{1,1,2} &$13146\PM 88       $&     -            & -               & -              &&$\uparrow    $&$\uparrow      $&$\uparrow    $& \\ 
 &$p\ON6        $&$0.050 \PM0.001    $&     -            & -               & -              &&$F\ON3816    $&$F\ON2111      $&$F\ON1567    $& \PR{dcm.dat} = \PR{jd14R112s.dat}                 \\
 &$R  \ON0.3605$ &                    &                  &                 &                &&\REJ          &\REJ            &\REJ          &                                                   \\ 
\hline  
2&\RModel{2,1,2} &$13421\PM20        $&$31957\pm 297    $& -               & -              && -            &$\uparrow      $&$\uparrow    $& \\ 
 &$p\ON9        $&$0.0585\PM0.0005   $&$0.079\PM0.001   $& -               & -              && -            &$F\ON36.0      $&$F\ON39.2    $& \PR{dcm.dat} = \PR{jd14R212s.dat}                  \\
\end{tabular}
\addtolength{\tabcolsep}{+0.18cm}
\label{Tablejd14}
\end{center}
\end{tiny}
\end{table*}

\begin{table*}
\caption{Comparison of long period search results. }
\begin{center}
  \begin{tiny}
\addtolength{\tabcolsep}{-0.14cm}
  \begin{tabular}{ccccccccccccccccc}
    \multicolumn{5}{c}{Table \ref{TableHJD}} & &
    \multicolumn{5}{c}{Table \ref{TableD}}   & &
    \multicolumn{5}{c}{Table \ref{TableJD}}   \\
    \multicolumn{5}{c}{\HJDdata} & &
    \multicolumn{5}{c}{\Ddata} & &
    \multicolumn{5}{c}{\JDdata}   \\
    \multicolumn{5}{c}{$1^{\mathrm{y}}$ signal=No, $1^{\mathrm{y}}$ window=Yes } & &
    \multicolumn{5}{c}{$1^{\mathrm{y}}$ signal=No, $1^{\mathrm{y}}$ window=No} & &
    \multicolumn{5}{c}{$1^{\mathrm{y}}$ signal=Yes, $1^{\mathrm{y}}$ window=Yes}   \\
    \cline{1-5}
    \cline{7-11}
    \cline{13-17}
    \multicolumn{2}{c}{Table \ref{Tablehjd13}: \threetest} & &
    \multicolumn{2}{c}{Table \ref{Tablehjd14}: \fourtest}  & & 
    \multicolumn{2}{c}{Table \ref{Tabled13}: \threetest}   & &
    \multicolumn{2}{c}{Table \ref{Tabled14}: \fourtest}    & &
    \multicolumn{2}{c}{Table \ref{Tablejd13}:  \threetest} & &
    \multicolumn{2}{c}{Table \ref{Tablejd14}:   \fourtest} \\
M &$P  \& A   \Dd  $&  & M &$P \&A  \Dd     $& & M &$P \& A \Dd     $& & M &$P\&A \Dd$& &M&$P \& A \Dd     $& & M&$P \& A \Dd$ \\  
    \cline{1-2}
    \cline{4-5}
    \cline{7-8}
    \cline{10-11}
    \cline{13-14}
    \cline{16-17}
3 &$32948 \pm 273       $&& 4&$33507\pm246           $&& 3&$33079\pm280         $&& 4&$33766\pm282         $&& 3&$32705\pm429         $&& 4&$33140\pm374    $\\
  &$0.084 \pm0.001      $&&  &$0.086\pm0.001         $&&  &$0.084\pm0.001       $&&  &$0.087\pm0.001       $&&  &$0.082\pm0.002       $&&  &$0.084\pm0.001  $\\
    \hline
3 &$13439\pm 23         $&& 4&$13537\pm28            $&& 3&$13449\pm31          $&& 4&$13547\pm16          $&& 3&$13438\pm32          $&& 4&$13534\pm24     $\\
\hline
\end{tabular}
\addtolength{\tabcolsep}{+0.14cm}
\end{tiny}
\end{center}
\label{TableComparison}
\end{table*}

\begin{table*}
  \caption{Short period search between $300^{\mathrm{d}}$ and $3000^{\mathrm{d}}$.
  Fours best periods for the eight signal residuals
  (\HJDdata: M=1-4, \Ddata: M=5-8, \JDdata: M=9-12), otherwise as in Table \ref{TableBest}. }
  \begin{center}
    \begin{tiny}
    \addtolength{\tabcolsep}{-0.08cm}
  \begin{tabular}{clccccccccc}
    \hline
    \multicolumn{11}{c}{\HJDdata}                                                                                                              \\
    \multicolumn{11}{c}{Eight signal residuals (\PR{file1} =\PR{hjd58R410sResiduals.dat})}                                                               \\
     &                & \multicolumn{4}{c}{Period analysis}                            &&\multicolumn{3}{c}{Fisher-test} &                   \\
    \cline{3-6} \cline{8-10}
  M  & Model          & $P_1 \& A_1 \Dd $& $P_2  \& A_2   \Dd  $& $P_3  \& A_3   \Dd  $&$P_4  \& A_4   \Dd  $&&\RModel{2,1,0}&\RModel{3,1,0}&\RModel{4,1,0}                                            \\
\hline
  1  & \RModel{1,1,0} &$616.4\pm 11     $&         -            &         -            &$                   $&&$\leftarrow$&$\uparrow   $&$\uparrow$& \\ 
     &$p=4           $&$0.0013\pm 0.0002$&         -            &         -            &$                   $&&$F=4.8     $&$F=4.2      $&$F=5.2   $& \PR{dcm.dat} = \PR{hjd14R110l.dat}               \\
     &$R=0.01462     $&                  &                      &                      &$                   $&&$Q_F=0.0026$&$Q_F=0.00038$&$Q_F=5.6\times10^{-7}$&                                      \\
    \hline
  2  & \RModel{2,1,0} &$365.5 \pm 10    $&$616.6\pm 14         $&         -            &$                   $&& -         &$\leftarrow  $&$\uparrow$& \\
\end{tabular}
  \end{tiny}
\end{center}
\addtolength{\tabcolsep}{+0.08cm}
\label{TableShort}
\end{table*}

\begin{table}
  \caption{$M_1$ and $M_2$ coefficients of $p(t)$.} 
\begin{center}
  \begin{tabular}{lcc}
    \hline
Model & $M_1$ & $M_2$       \\
    \hline
 Table \ref{Tablehjd14}: M=1  & $-0.930 \pm 0.007$ & $0.209 \pm 0.003$ \\
 Table \ref{Tablehjd14}: M=2  & $-0.646 \pm 0.006$ & $0.089 \pm 0.002$ \\
 Table \ref{Tablehjd14}: M=3  & $-0.635 \pm 0.006$ & $0.085 \pm 0.002$ \\
 Table \ref{Tablehjd14}: M=4  & $-0.627 \pm 0.005$ & $0.082 \pm 0.002$ \\
\hline
  \end{tabular}
  \end{center}
\label{TableM}
\end{table}

\end{document}